\documentstyle[aps,prl,epsf]{revtex} 
\begin{document}

\twocolumn[\hsize\textwidth\columnwidth\hsize  
\csname @twocolumnfalse\endcsname              
\vskip -0.8cm
{\hfill ORNL-CTP-96-07 (hep-ph/9607285)}
\vskip 0.3cm

\title{ Suppression of $J/\psi$ and $\psi'$ Production in High-Energy
Pb on Pb Collisions}

\author{ Cheuk-Yin Wong }
\address{ Oak Ridge National Laboratory, Oak Ridge, TN 37831}


\maketitle

\begin{abstract}

The anomalous $J/\psi$ suppression in Pb-Pb collisions at 158A GeV
observed recently by NA50 can be explained as due to the transition to
a new phase of strong $J/\psi$ absorption, which sets in when the
local energy density exceeds about 3.4 GeV/fm$^3$.

\end{abstract}


   ]                               

\narrowtext


High-energy heavy-ion collisions have become the focus of intense
research because of the possibility of producing a quark-gluon plasma
during such collisions \cite{QM93,Won94}. The suppression of $J/\psi$
and $\psi'$ production has been suggested to probe the screening
between a charm quark and a charm antiquark in the plasma
\cite{Mat86,Gup92}.  While $J/\psi$ and $\psi'$ suppression has been
observed \cite{Bag89,Ald91,Lou95,Bag95}, the phenomenon can be
explained by absorption models without assuming the occurrence of the
plasma \cite{Ger88,Vog91,Won96a,Kha96,Huf96,Woncw96}.

Recently NA50 observed that the $J/\psi$ and $\psi'$ production is
anomalously suppressed for Pb-Pb collisions at 158A GeV \cite{Gon96}.
The observation led to a flurry of activities.  While the present
author and others presented theoretical findings at the Quark Matter
'96 Meeting in May, 1996 \cite{Won96qm,Kha96qm,Bla96,Gav96}, other
theoretical studies have since been put forth \cite{Cap96,Cas96}.  A
central question is whether the anomalous suppression in Pb-Pb
collisions arises from the absorption by comovers (produced hadrons),
as proposed by \cite{Gav96,Cap96,Cas96}, or from the occurrence of a
new phase of strong $J/\psi$ absorbing matter (possibly a quark-gluon
plasma), as suggested in \cite{Won96qm,Kha96qm,Bla96}.  The anomalous
suppression of $\psi'$ in Pb-Pb \cite{Gon96} and S-U
collisions\cite{Lou95,Bag95} is another question which needs to be
addressed in the context of any possible occurrence of the new phase.

The study of these two questions requires the examination of
absorption by soft particles not in the new phase, which is beyond the
scope of the schematic model of \cite{Bla96}.  Previously, a
microscopic absorption model (MAM) was proposed which allows one to
study the absorption by soft particles \cite{Won96a}.  We shall use
the MAM model to examine the two central questions outlined above.  We
show that contrary to the conclusions of \cite{Gav96,Cap96,Cas96},
absorption by produced soft particles cannot explain the anomalous
$J/\psi$ absorption in Pb-Pb collisions; that a class of models with
the feature of the occurrence of a new phase of strong absorption can
explain the complete set of $J/\psi$ data; and that $\psi'$ is already
anomalously absorbed by produced soft particle not in the new phase,
and the transition to the new phase leads only to a small increase in
$\psi'$ absorption in Pb-Pb collisions.

We envisage that a produced (quasi-bound) $c\bar c$ pair evolving into
$J/\psi$ and $\psi'$ states will collide with two types of particles.
Collisions with baryons occur at high relative energies, and
constitute the hard component of the absorption process.  The
absorption cross sections $\sigma_{\rm abs}(\psi N)$ and $\sigma_{\rm
abs}(\psi' N)$ at high energies are approximately equal empirically
\cite{Bri83}, which can be understood by the Glauber picture of
hadron-hadron collision \cite{Won96a}.  The collisions of the $c\bar
c$ with produced gluons or produced hadrons, which constitutes the
soft component of absorption, occur at low relative energies of about
200 MeV, the temperature of produced particles.  The breakup threshold
is about 640 MeV for $J/\psi$ and only about 50 MeV for $\psi'$.  One
expects the soft component to be small for $J/\psi$ but large for
$\psi'$.  This is borne out by the data in Fig. 1 where ${\cal
B}\sigma/AB$ is plotted (in logarithmic scale) as a function of
$\eta=A^{1/3}(A-1)/A+B^{1/3}(B-1)/B$, which is approximately
proportional to the average path length passing through nuclei A and B
\cite{Won94}.  The absorption factor for the hard component is
$\exp\{-{\rm constant}\times\eta\}$.  The absorption factor for the
soft component is similarly $\exp\{- {\rm constant}'\times\eta\}$
because the density of the produced soft particles is proportional to
the longitudinal path length passing through nuclei A and B (see pages
374-377 of \cite{Won94}).  $p$-A collisions involve the hard component
while A-B collisions involve both the hard and soft components
\cite{Won96a}.  The magnitude of the soft component is indicated by a
gap and a slope change between the A-B line and the $p$-A line in
Fig. 1.  A large soft component\break
\epsfxsize=200pt
\includegraphics{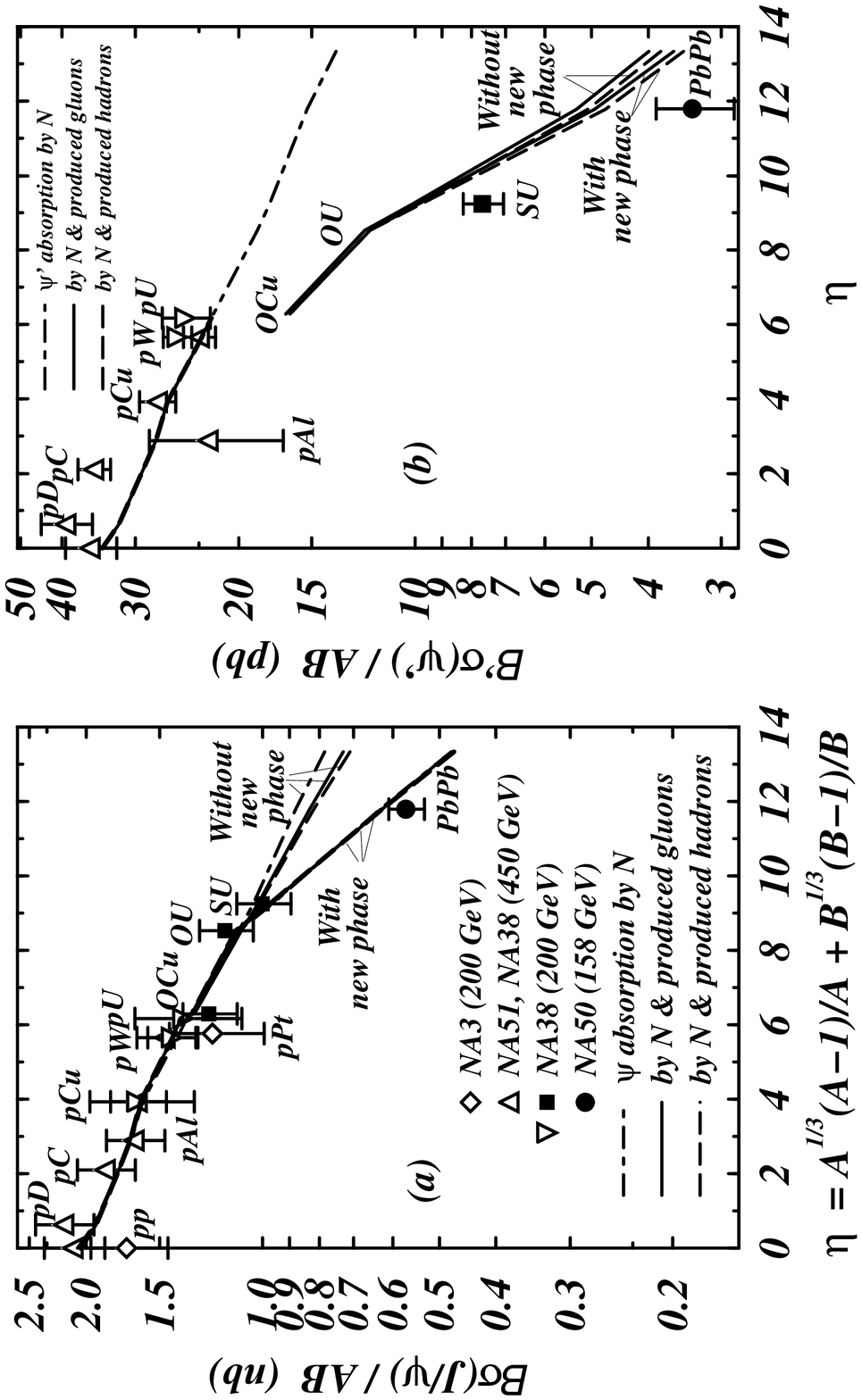}
\vskip 5.5cm
\begin{minipage}[t]{8cm}
\noindent \bf Fig.1.  \rm
{ ($a$) \protect${\cal B}\sigma_{J/\psi}^{AB}/AB\protect$ and
($b$) \protect${\cal B}'\sigma_{\psi'}^{AB}/AB\protect$ as a function
of $\eta$.  Data are from NA3 \protect\cite{Bad83}, NA51
\protect\cite{Bal94}, NA38 \protect\cite{Bag89,Lou95,Bag95}, and NA50
\protect\cite{Gon96}.  }
\end{minipage}
\vskip 4truemm
\noindent 
for $\psi'$ absorption is indicated in
Fig.\ 1$b$ by a large gap between the S-U point and the $p$-A line. A
very small soft component for $J/\psi$ absorption is indicated in
Fig. 1$a$, as there is almost no gap and no slope change between the
$p$-A line and the A-B line joining the O-Cu, O-U, and S-U points.
The A-B line passing the O-Cu, O-U and S-U points is much above the
Pb-Pb point in Fig.\ 1$a$, indicating that the soft component
constrained to explain the data of O-Cu, O-S and S-U cannot explain
the anomalous $J/\psi$ suppression in Pb-Pb collisions. A new source
of absorption is suggested.

We shall further confirm the above observations using the microscopic
absorption model (MAM).  Adopting a row-on-row picture in the
center-of-mass system and following straight-line space-time
trajectories of the $c\bar c$'s, the baryons, and the centers of the
fireballs of produced soft particles, the differential cross section
for $J/\psi$ production in an $A$-$B$ collision is \cite{Won96a}
$$
{ d \sigma_{{}_{J/\psi}}^{{}^{AB}} (\bbox{b}) \over \sigma_{{}_{J/\psi
}}^{{}^{NN}}~d\bbox{b} } = \!\! \int \!\! { d{\bbox{b}}_{{}_A} \over
\sigma_{\rm abs}^2(\psi N) } \biggl \{ 1 -\biggl [ 1-
T_{{}_A}(\bbox{b}_{{}_A}) \sigma_{\rm abs}(\psi N) \biggr ] ^A \biggr
\}
$$
\begin{eqnarray}
\label{eq:fin}
~~~~~~~\times \biggl \{ 1 -\biggl [ 1- T_{{}_B}(\bbox{b}-\bbox{b}_{{}_A})
\sigma_{\rm abs}(\psi N) \biggr ] ^B \biggr \} F(\bbox{b}_A)\,,
\end{eqnarray}
where $T_A(\bbox{b}_A)$ is the thickness function of nucleus $A$, 
and the soft particle absorption factor $F(\bbox{b}_A)$ is 
\begin{eqnarray}
\label{eq:fb}
F(\bbox{b}_A) \!= \!\sum_{n=1}^{N_<}  \!{ a(n)\over N_> N_<} 
\!\sum_{j=1}^n
\!\exp\{- \theta \!\!\!\! \sum_{i=1, i\ne j}^n 
\!\!\!( k_{\psi g} t_{ij}^g +
k_{ \psi h} t_{ij}^h ) \} \,.
\end{eqnarray}
Here, $N_>(\bbox{b}_A)$ and $N_<(\bbox{b}_A)$ are the greater and the
lesser of the (rounded-off) nucleon numbers
$AT_A(\bbox{b}_A)\sigma_{in}$ and
$BT_B(\bbox{b}-\bbox{b}_A)\sigma_{in}$ in the row at $\bbox{b}_A$ with
an $NN$ inelastic cross section $\sigma_{in}$, $a(n)=2 {\rm~~for~~}
n=1,2,...,N_<-1 $, and $ a(N_<)=N_>-N_<+1$.  The quantities
$\sigma_{\rm abs}(\psi m)$, the average relative velocity $v_m$, and
the average number density $\rho_m$ per $NN$ collision always come
together so that we can use the rate constant $k_{\psi m}$ ($m=g, h$)
to represent their product.  The interaction time $t_{ij}^{g}$ (or
$t_{ij}^{h}$) is the time for a $J/\psi$ produced in collision $j$ to
overlap with the center of the fireball of gluons (or hadrons)
produced in collision $i$ at the same spatial point. They can be
determined from $c\bar c$, $g$, $h$ production time $t_{c\bar c}$,
$t_g$, $t_h$, and the freezeout time $t_f$. The function $ \theta$ is
zero if $A=1$ or $B=1$, and is 1 otherwise.  The expressions for
$\psi'$ production can be obtained from Eqs.\
(\ref{eq:fin}-\ref{eq:fb}) above by changing $\psi$ into $\psi'$.

We first study ${\cal B}\sigma/AB$ data without the Pb-Pb points and
compare them with theoretical models, including the possibility of
absorption by soft particles: (A) absorption by baryons only, as in
\cite{Ger88}; (B) by baryons and produced soft gluons, as in
\cite{Won96a}; and (C) by baryons and produced soft hadrons (similar
to \cite{Vog91} but differing in details and methods of evaluation).
Following \cite{Won96a}, we use $\sigma_{\rm abs}(\psi' N)=
\sigma_{\rm abs}(\psi N)$ and search for $\sigma_{\rm abs}(\psi N)$ as
well as $k_{\psi g}$ and $k_{\psi' g}$ in Model B, and $k_{\psi h}$
and $k_{\psi' h}$ in Model C, fixing the time constants to have the
plausible values $t_g=0.1 $, $t_h=1.2$, $t_f=3$, and $t_{c\bar
c}=0.06$ (in units of fm/c).  Different time constants will modify
inversely the rate constants but will not affect greatly the product
of the rate constants and their corresponding average interaction
times.  We take a Gaussian nuclear density for $A<40$ and a
Woods-Saxon density for $A\ge 40$.  The curves marked ``without new
phase'' in Fig. 1 are calculated with the parameters $\sigma_{\rm
abs}(\psi N)=6.94$ mb in Model A, and $\sigma_{\rm abs}(\psi N)=6.36 $
mb in Model B and C.  In addition, $k_{\psi g}=0.0956 $ c/fm and
$k_{\psi' g}=3 $ c/fm for Model B, and $k_{\psi h}=0.0493$ c/fm
$k_{\psi' h}=3$ c/fm for Model C.  As $k_{\psi g} << k_{\psi' g}$ in
Model B and $k_{\psi h} << k_{\psi' h}$ in Model C, we confirm that
the soft absorption component is small for $J/\psi$ but large for
$\psi'$.

When we include the Pb-Pb data point, no MAM calculations with a hard
and a soft absorption component can simultaneously describe the whole
set of $J/\psi$ data.  This suggests that there is a transition to a
new phase of strong $J/\psi$ absorption, when the local energy density
exceeds a certain threshold.  One can extend the MAM model to describe
this transition.  The energy density is approximately proportional to
the number of collisions which has taken place at that point up to
that time.  We postulate that soft particles make a transition to a
new phase of strong $J/\psi$ absorption if there have been $N_c$ or
more baryon-baryon collisions at that point at time $t_x$.  The
quantity $k_{\psi g} t_{ij}^g + k_{\psi h} t_{ij}^h$ in Eq.\ (2)
becomes $k_{\psi g} t_{ij}^g + k_{\psi h} t_{ij}^h + k_{\psi x}
t_{ij}^x$, where the new rate constant $k_{\psi x}$ describes the
absorption of $J/\psi$ by the produced soft matter in the new phase.
Here, $t_{ij}^x=t_n+t_h-t_x$ where $t_n$ is the last $NN$ collision
time at that point and $t_{ij}^x$ is the time for a $J/\psi$ produced
in collision $j$ to overlap with the center of the fireball of
absorbing soft particles produced in $i$ in the form of the new phase,
before hadronization takes place.  We vary $N_c$ and $k_{\psi x}$.
Baryons passing through the spatial region of the new phase may also
become deconfined and may alter their $\psi$-$N$ absorption cross
section.  Accordingly, we also vary the effective absorption cross
section, $\sigma_{\rm abs}^x(\psi N)$, for $\psi$-$N$ interactions in
the row in which there is a transition to the new phase, while
$\sigma_{\rm abs}(\psi N)$ remain unchanged in other rows.  The curves
marked ``with new phase'' in Fig. 1 are obtained with the parameters
$N_c=4$, $k_{\psi x}=1$ c/fm, and $\sigma_{\rm abs}^x(\psi N)=14$ mb.
The location where the slopes of the curves changes sharply is
sensitive to $N_c$.  The absorption of $J/\psi$ saturates for $k_{\psi
x} \ge 1$ c/fm for which the absorption is still slightly insufficient
to account for the total $J/\psi$ absorption; additional absorption
with $\sigma_{\rm abs}^x(\psi N)=14$ mb is needed to give Pb-Pb result
to agree with experiment.  The results for the cases with the new
phase agree with the whole set of $J/\psi$ data.  We note that
$k_{\psi x} >> k_{\psi g}, k_{\psi h}$.

The above MAM results can be expanded to obtain $J/\psi$ and $\psi'$
yields as a function of impact parameter $b$.  The theoretical ratio
${\cal B}\sigma_{J/\psi}^{AB}/\sigma^{AB}(DY)_{2.9-4.5}$ for Drell-Yan
cross section in the interval $2.9<M_{\mu^+\mu^-}<4.5$ GeV can then be
expressed in terms of ${\cal
B}\sigma_{J/\psi}^{pp}/\sigma^{pp}(DY)_{2.9-4.5}$ for $pp$ collisions,
which was determined by NA51 to be 44$\pm3$ \cite{Gon96}.  NA38 and
NA50 used a geometrical model to obtain a relation between $E_T$ and
$b$ \cite{Lou95,Gon96}, which can be used to transform the MAM results
from a function of $b$ to a function of $E_T$.  In Fig.\ 2$a$ we show
theoretical ${\cal B}\sigma_{J/\psi}^{AB}/ \sigma^{AB}(DY)_{2.9-4.5}$ as a
function of $E_{{}_T}$ for S-U which agree with experiment within
5-10\%, whether we assume a new phase or not.  On the other hand, good
agreement with Pb-Pb data is obtained only when we allow for the
transition to a new phase of strong $J/\psi$ absorption (Fig.\ 2$b$).

\vspace*{-1.1cm}
\hskip -1.0cm
\epsfxsize=200pt
\includegraphics{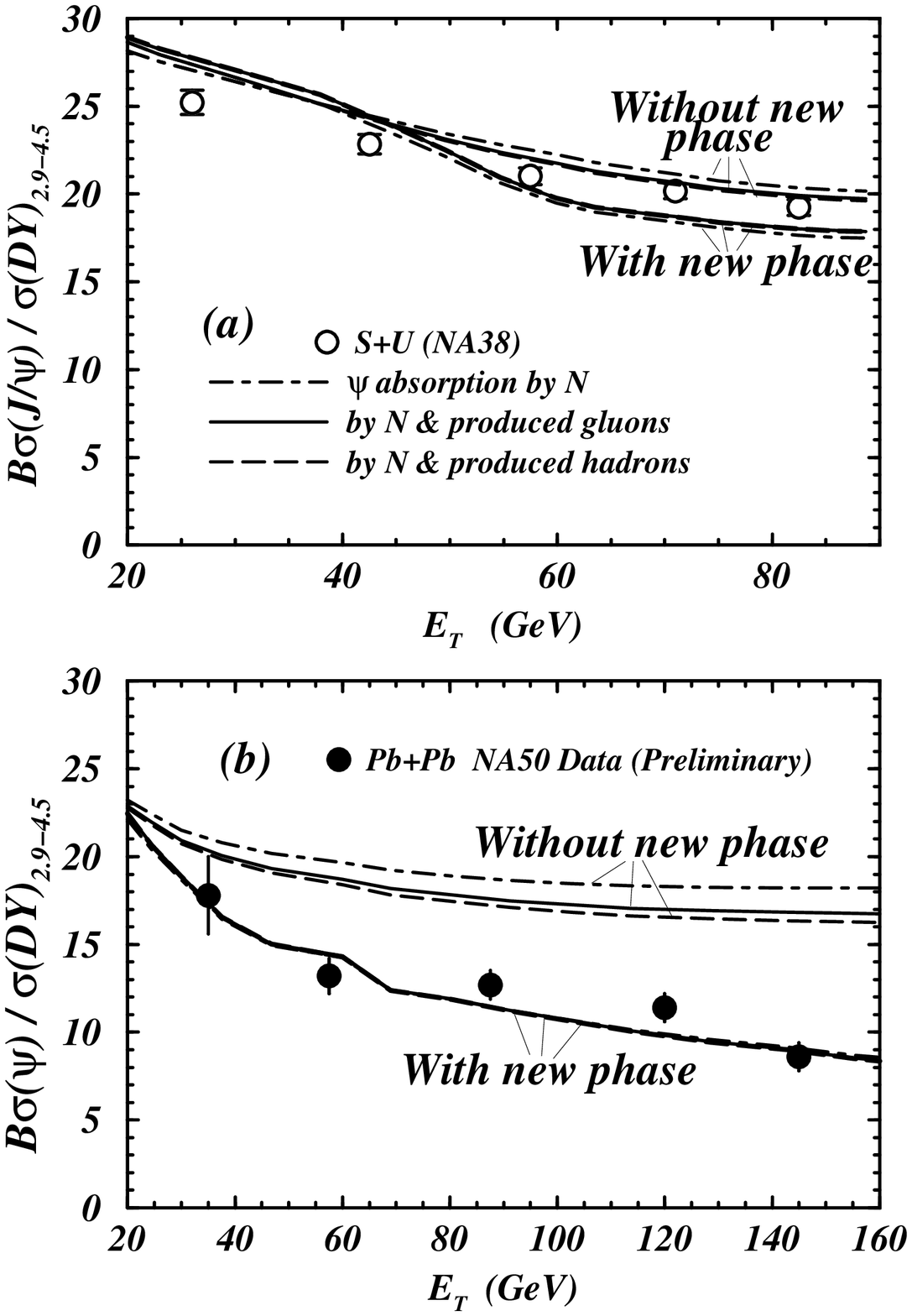}
\vskip 9.7cm
\begin{minipage}[t]{8cm}
\noindent \bf Fig.2.  \rm
{ \protect${\cal B}
\protect\sigma_{J/\psi}^{AB}/\protect\sigma(DY)_{2.9-4.5} \protect$ as
a function of $E_{{}_T}$ for ($a$) S-U collisions, and ($b$) Pb-Pb
collisions.  Data are from NA38 \protect\cite{Bag95} and NA50
\protect\cite{Gon96}.  }
\end{minipage}
\vskip 4truemm
\noindent

The results in Fig.\ 1$b$ indicate that $\psi'$ is already
anomalously absorbed by soft particles in S-U and Pb-Pb collisions,
even without the new phase.  The occurrence of the new phase with
$k_{\psi' x}=3$ c/fm leads only to a small additional $\psi'$
absorption in Pb-Pb collisions (Fig.\ 1$b$).  We can calculate ${\cal
B}'\sigma (\psi')/{\cal B}\sigma (J/\psi)$ and compare with
experiment.  Fig. 3$b$ shows that the experimental data for Pb-Pb
collisions is consistent with the assumption of transition to the new
phase.  For S-U collisions, the theoretical results agree with
experimental data for small and moderate values of $E_T$, but deviate
from experimental data for large $E_T$ (Fig.\ 3$b$).  As large
transverse energies involve greater weights for the major axis of the
U nucleus to lie along the beam direction, the deviation of ${\cal
B}'\sigma (\psi')/{\cal B}\sigma (J/\psi)$ in S-U collisions at high
$E_T$ may be a deformation effect which can be tested experimentally
by studying S-Pb collisions.  The deformation effect can be utilized
to study matter in the new phase by focusing on high $E_T$ events in
U-U collisions.

\vspace*{-0.9cm}
\hskip -1.0cm
\epsfxsize=200pt
\includegraphics{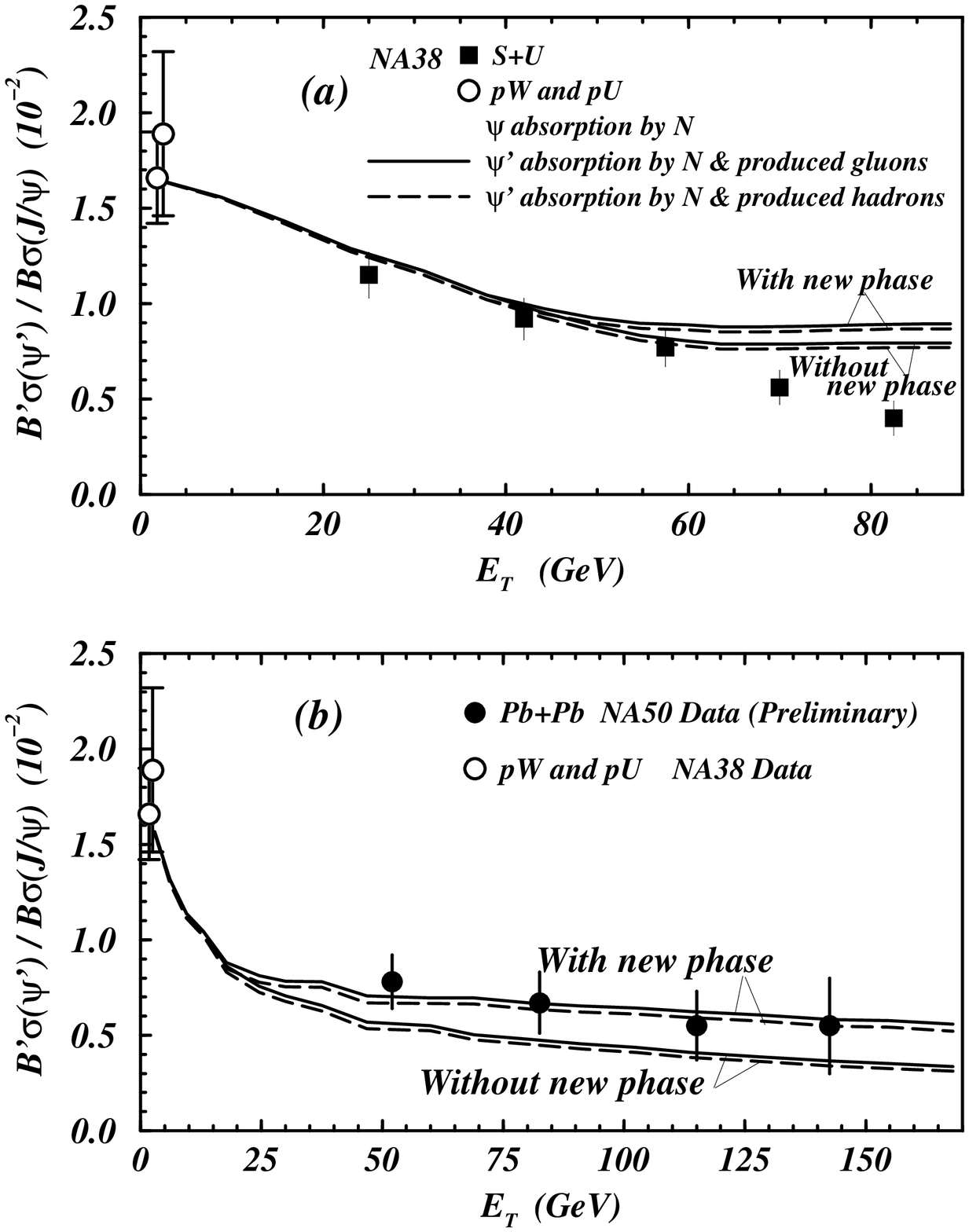}
\vskip 9.7cm
\begin{minipage}[t]{8cm}
\noindent \bf Fig.3.  \rm
{ ${\protect {\cal B}}'\sigma(\psi')/ {\protect {\cal
B}}\sigma(\psi)$ as a function of $E_T$ for (a) S-U collisions and (b)
Pb-Pb collisions.  The data for $p$-W and $p$-U collisions are also
included.  Data are from NA38 \protect\cite{Bag95,Lou95} and NA50
\protect\cite{Gon96}.  }
\end{minipage}
\vskip 4truemm
\noindent

What is the threshold energy density for the transition?  We consider
a row with a transverse area $\sigma_{in}$ and with collision points
separated longitudinally by $d/\gamma$, where $d=2.46$ fm is the
internucleon spacing and $\gamma=\sqrt{s}/2m_{\rm nucl}=9.2$ is the
Lorentz contraction factor at the Pb-Pb collision energy.  Consider
also $N_c$ number of $NN$ collisions at each collision point. Each
$NN$ collision leads to a fireball of soft particles with
$dn^{NN}/dy\sim 1.9 $ at the Pb-Pb collision energy \cite{Tho77}.  At
the time $t=d/\gamma$ after these $N_c$ collisions, the produced soft
particles leaving the fireballs at one collision point to the adjacent
collision point will be compensated by soft particles arriving from
the fireballs of the adjacent collision point, and a steady-state
initial energy density is reached after $t=d/\gamma$ (before
longitudinal expansion), with the energy density given approximately
by $N_c (dn^{NN}/dy) m_t /(\sigma_{in}d/\gamma)$, where $m_t=0.35$ GeV
is the pion transverse mass. Hence, for $N_c=4$ from the present
study, the threshold energy density for the new phase is $\epsilon_c
\sim 3.4$ GeV/fm$^3$, which is close to the quark-gluon plasma energy
density, $\epsilon_c \sim 4.2$ GeV/fm$^3$, calculated from the lattice
gauge result $\epsilon_c/T_c^4 \sim 20$ \cite{Blu95} with $T_c \sim
0.2$ GeV.  Therefore, it is interesting to speculate whether the new
phase of strong $J/\psi$ absorption may be the quark-gluon plasma.  In
the equilibrated or non-equilibrated quark-gluon plasm, $J/\psi$
production is expected to be greatly suppressed \cite{Mat86,Xu95}.
Furthermore, when baryons pass through the region of the new phase,
the baryon matter may be deconfined.  The total cross section between
a $c\bar c$ system and a baryon system is substantially enhanced when
the quarks in the baryon are deconfined \cite{Won96b}.

Our results for the cases with soft particle absorption differ
qualitatively from those of \cite{Gav96,Cap96,Cas96}.  We would like
to mention some of the reasons for the differences.  In
Refs. \cite{Gav96,Cap96}, the soft particle absorption factor is
approximately the form $\exp\{-{\rm constant}\times E_T(b)\}$.  This
is obtained in \cite{Gav96} by assuming the density of produced soft
particles to be $n(b)={\rm constant}\times E_T(b)$.  Ref.\
\cite{Gav96} assumes $n(b)={\rm constant}\times G(b)E_T(b)$ and finds
$G(b)$ to be independent of $b$.  However, $E_T(b)$ contains
contributions coming from the whole volume of $NN$ collisions of the
participant nucleons, which decreases significantly as $b$ increases.
When this volume ${ V}(b)$ (in the C.M. system) is taken into account,
the proper relation should be $n(b) = {\rm constant} \times E_T(b)/{
V}(b) $.  The additional ${ V}(b)$-type dependence in $n(b)$ will
modify the results of \cite{Gav96,Cap96}.  Fig.\ 2 of the numerical
cascade model of \cite{Cas96} gives a resultant $J/\psi$ absorption
factor for $J/\psi$ absorption due to baryons in S-W collisions to be
smaller than that in $p$-W collision at the same impact parameter of 2
fm.  The authors of \cite{Cas96} should check whether these results
are consistent with intuitive understanding and analytic results for
absorption due to baryons.  Clearly, much work remains to be done to
resolve the differences.

\acknowledgements

The author would like to thank M.\ Gonin, C.\ Louren\c co and C.\ W.\
Wong for helpful communications.  This research was supported by the
Division of Nuclear Physics, U.S. D.O.E.  under Contract
DE-AC05-96OR22464 managed by Lockheed Martin Energy Research Corp.

\end{document}